\documentclass[twocolumn,showpacs,preprintnumbers,amsmath,amssymb]{revtex4}

\usepackage{graphicx}
\usepackage{dcolumn}
\usepackage{bm}

\begin{document}

\title{Maximal planar networks with large clustering coefficient and power-law degree distribution}
\author{Tao Zhou$^{1,2}$}
\author{Gang Yan$^{2}$}
\author{Bing-Hong Wang$^{1}$}
\email{bhwang@ustc.edu.cn,Fax: +86-551-3603574.}
\affiliation{%
$^{1}$Nonlinear Science Center and Department of Modern Physics,\\
University of Science and Technology of China, \\
Hefei Anhui, 230026, PR China \\
$^{2}$Department of Electronic Science and Technology,\\
University of Science and Technology of China, \\ Hefei Anhui,
230026, PR China
}%

\date{\today}

\begin{abstract}
In this article, we propose a simple rule that generates
scale-free networks with very large clustering coefficient and
very small average distance. These networks are called {\bf Random
Apollonian Networks}(RAN) as they can be considered as a variation
of Apollonian networks. We obtain the analytic results of
power-law exponent $\gamma =3$ and clustering coefficient
$C=\frac{46}{3}-36\texttt{ln}\frac{3}{2}\approx 0.74$, which agree
very well with the simulation results. We prove that the
increasing tendency of average distance of RAN is a little slower
than the logarithm of the number of nodes in RAN. Since most
real-life networks are both scale-free and small-world networks,
RAN may perform well in mimicking the reality. The RAN possess
hierarchical structure as $C(k)\sim k^{-1}$ that in accord with
the observations of many real-life networks. In addition, we prove
that RAN are maximal planar networks, which are of particular
practicability for layout of printed circuits and so on. The
percolation and epidemic spreading process are also studies and
the comparison between RAN and Barab\'{a}si-Albert(BA) as well as
Newman-Watts(NW) networks are shown. We find that, when the
network order $N$(the total number of nodes) is relatively
small(as $N\sim 10^4$), the performance of RAN under intentional
attack is not sensitive to $N$, while that of BA networks is much
affected by $N$. And the diseases spread slower in RAN than BA
networks during the outbreaks, indicating that the large
clustering coefficient may slower the spreading velocity
especially in the outbreaks.
\end{abstract}

\pacs{89.75.Hc, 64.60.Ak, 84.35.+i, 05.40.-a, 05.50+q, 87.18.Sn}

\maketitle

\section{Introduction}

Many social, biological, and communication systems can be properly
described as complex networks with nodes representing individuals
or organizations and edges mimicking the interactions among
them\cite{Albert2002,Dorogovtsev2002,Newman2003,Newman2000,Hayes2000a,Hayes2000b,Strogatz2001,Wang2002,Wang2003,Evans2004,Amaral2004,Barabasi2002,Dorogovtsev2003a,Watts1999a}.
Examples are numerous: these include the
Internet\cite{Faloutsos1999,Pastor2001,Caldarell2000}, the World
Wide
Web\cite{Huberman2001,Barabasi2000,Flake2002,Broder2000,Albert1999},
social networks of acquaintance or other relations between
individuals\cite{Scott2000,Wasserman1994,Liljeros2001,Potterat2002,Morris1997,Newman2001a,Newman2001b,Fan2004,Li2004,Hu2004a,Hu2004b},
metabolic
networks\cite{Jeong2000,Podani2001,Fell2000,Wagner2001a,Stelling2002},
food
webs\cite{Pimm2002,Montoya2002,Sole2001,Camacho2002,Williams2002,Dunne2002a,Dunne2002b}
and many
others\cite{Redner1998,Amaral2000,He2004,Xu2004,Latora2002,Sen2003,Maritan1996,Ferrer2001,Sporns2002,Banavar1999,West1997,West1999}.
The ubiquity of complex networks inspires scientists to construct
a general model. In the past 200 years, the study of topological
structures of the networks used to model the interconnection
systems has gone through three stages. For over a century, there
is an implicit assumption that the interaction patterns among the
individuals can be embedded onto a regular structure such as
Euclidean lattices, hypercube networks, and so
on\cite{Xu2001,Xu2003,Bond1976,Bollobas1998}. Since late 1950s
mathematicians began to use random graphs to describe the
interconnections, this is the second
stage\cite{Erdos1947,Solomonoff1951,Erdos1959,Erdos1960,Erdos1961,Karonski1982,Bollobas1985,Janson1999}.
In the past few years, with the computerization of data
acquisition process and the availability of high computing powers,
scientists have found that most real-life networks are neither
completely regular nor completely random. The results of many
experiments and statistical analyses indicate that the networks in
various fields have some common characteristics, the most
important of which are called small-world
effect\cite{Milgram1967,Watts1998,Watts1999b,Newman1999a,Monasson1999,Zhu2004}
and scale-free
property\cite{Barabasi1999a,Barabasi1999b,Price1965,Price1976}.

In a network, the distance between two nodes is defined as the
number of edges along the shortest path connecting them. The
average distance $L$ of the network, then, is defined as the mean
distance between two nodes, averaged over all pairs of nodes. The
average distance is one of the most important parameters to
measure the efficiency of communication networks. For instance, in
a store-forward computer network, probably the most useful
measurement of its performance is the transmission delay (or time
delay) encountered by a message travelling through the network
from its source to destination; this is approximately proportional
to the number of edges a message must pass through. Thus the
average distance plays a significant role in measuring the
transmission delay. The number of the edges incident from a node
$x$ is called the degree of $x$, denoted by $k(x)$. Obviously,
through the $k(x)$ edges, there are $k(x)$ nodes that are
correlated with $x$; these are called the neighbor-set of $x$, and
denoted by $A(x)$. The clustering coefficient $C(x)$ of node $x$
is the ratio between the number of edges among $A(x)$ and the
total possible number, the clustering coefficient $C$ of the whole
network is the average of $C(x)$ over all $x$. Experiments
indicate that most real-life networks have much smaller average
distance (as $L\sim\ln N$ where $N$ is the number of nodes in the
network) than the completely regular networks and have much
greater clustering coefficient than those of the completely random
networks. Therefore they should not be treated as either
completely regular or random networks. The recognition of
small-world effect involves the two factors mentioned above: a
network is called a small-world network as long as it has small
average distance and great clustering coefficient. Another
important characteristic in real-life networks is the power-law
degree distribution, that is $p(k)\propto k^{-\gamma}$, where $k$
is the degree and $p(k)$ is the probability density function for
the degree distribution. $\gamma$ is called the power-law
exponent, and usually between 2 and 3 in real-life
networks\cite{Albert2002,Dorogovtsev2002,Newman2003}. This
power-law distribution falls off much more gradually than an
exponential one of a completely random network, allowing for a few
nodes of very large degree to exist. Networks with power-law
degree distribution are referred to as scale-free networks,
although one can and usually does have scales present in other
network properties.

\begin{figure}
\scalebox{0.5}[0.5]{\includegraphics{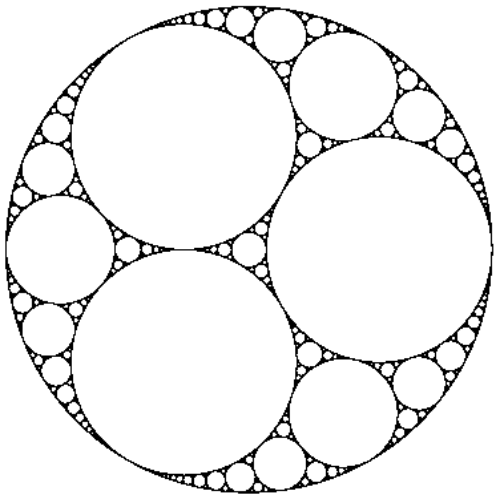}} \caption{An
Apollonian packing of disks within a circle\cite{Doye2004}.}
\end{figure}

From 1998 much attention has been focused on how to model complex
network. One of the most well-known models is Watts and Strogatz's
small-world network (WS network), which can be constructed by
starting with a regular network and randomly moving one endpoint
of each edge with probability $p$ \cite{Watts1998}. Another
popular model was proposed independently by
Monasson\cite{Monasson1999} and by Newman and Watts(NW
Networks)\cite{Newman1999a}, where no edges are rewired. Some
variations of the small-world model have been proposed. Several
authors have studied the model in dimension higher than one and
obtained similar results to the one dimension
case\cite{Newman1999a,Menezes2000,Moukarzel1999,Newman1999b,Ozana2001}.
Another kind of models in which shortcuts preferentially join
nodes that are close together on the underlying lattice also have
been studied\cite{Jespersen2000,Kleinberg2000,Sen2001}. Very
recently, Zhu et. al. proposed a so-called directed dynamical
small-world model, in which the network structure is affected by
the processes upon the network\cite{Zhu2004}.

Another significant model is Barab\'{a}si and Albert's scale-free
network model (BA network)\cite{Barabasi1999a,Barabasi1999b},
which is very much similar to Price's\cite{Price1965,Price1976}.
The BA model suggests that two main ingredients of
self-organization of a network in a scale-free structure are
growth and preferential attachment. These point to the facts that
most networks grow continuously  by adding new nodes, which are
preferentially attached to existing nodes with large number of
neighbors. The subsequent researches on various processes taking
place upon complex networks, such as
percolation\cite{Newman1999b,Albert2000a,Moore2000,Callaway2000,Cohen2000,Cohen2001,Dorogovtsev2001a,Schwartz2002,Sarshar2004,Jesch2004,Palla2004,Dorogovtsev2003b,Dorogovtsev2004},
epidemic
processes\cite{Pastor-Satorras2001a,Pastor-Satorras2001b,Boguna2002,Boguna2003,May2001,Pastor-Satorras2002,Barthelemy2004,Yan2004},
cascade
processes\cite{Holme2002a,Holme2002b,Moreno2002,Moreno2003,Motter2002,Motter2004,God2003,Lee2004,Zhou2004a}
and so on, indicate that the scale-free degree distribution plays
the most crucial role rather than small-world effect. Therefore,
in the recent two or three years, the study of modelling complex
networks focuses on revealing the underlying mechanism of
power-law degree distribution. Roughly, these models for
scale-free networks can be classified into 3 main scenarios. The
first one is related to the models of human
behavior\cite{Simon1955,Bornholdt2001} and was introduced in a
network version under the name ``preferential attachment"
mentioned
above\cite{Barabasi1999a,Barabasi1999b,Krapivsky2000,Krapivsky2001,Dorogovtsev2000a,Dorogovtsev2000b,Dorogovtsev2001b,Albert2000b,Tadic2001,Tadic2002,Bianconi2001a,Bianconi2001b,Yuan2004,Deng2004,LiCG2004}.
The second class of models is where a scale-free distribution
appears as a result of a balance between a modelled tendency to
form hubs against an entropic pressure towards a random network
with an exponential degree
distribution\cite{Capocci2001,Berg2002,Baiesi2003}. The third one
is the self-organized models that lead to power-law degree
distribution\cite{Kim2004,Yan2004b,Jain1998,Jain2001,Sole2002,Vazquez2003,Wagner2001b}.

In the recent several months, a few authors have demonstrated the
use of pure mathematical objects and methods to construct
scale-free networks. The first interesting instance is the
so-called integer networks\cite{Zhou2004b}, in which the nodes
represent integers and two nodes $x$ and $y$ are linked by an edge
if and only if $x$ is divided exactly by $y$ or $y$ is divided
exactly by $x$, where $x$ and $y$ are nonzero integers. Zhou and
Wang et al have studied the statistical properties of these
networks and demonstrated that they are scale-free networks of
large clustering coefficient. Another significant instance is {\bf
Apollonian Networks}(AN) introduced by Andrade et
al\cite{Andrade2004}. In our opinion, Apollonian networks may be
not the networks of best performance, but assuredly the most
beautiful ones we have ever seen. Another related work is owed to
Dorogovtsev and Mendes et. al., in which the deterministic
networks, named {\bf pseudofractals}, are obtained by random
attachment aiming at
edges\cite{Dorogovtsev2001c,Dorogovtsev2002b}.

In this article, we propose a simple rule that generates
scale-free networks with very large clustering coefficient and
very small average distance. These networks are called {\bf Random
Apollonian Networks}(RAN), since they can be considered as a
variation of Apollonian networks\cite{Zhou2004c}. We discuss the
difference between RAN and AN in detail(in addition, the
difference between RAN and BA networks), and demonstrates that RAN
perform much better than BA networks in some aspects.

This article is organized as follow: In section 2, the Apollonian
networks is introduced, including its fascinating properties and
its shortcomings. In section 3, the rule that generates RAN is
described in detail. In section 4, we give both the simulation and
analytic results about the statistical characteristics of RAN,
including the scale-free property and the small-world effect,
where the detailed analytic processes are shown in appendices. In
section 5, we prove that RAN are the maximal planar networks. In
section 6, The percolation and epidemic spreading process are also
studies and the comparison between RAN and BA networks as well as
NW networks are shown. Finally, we sum up this article and discuss
the relevance of RAN to the real world in section 7.

\section{Brief introduction of Apollonian networks}

\begin{figure}
\scalebox{0.7}[0.7]{\includegraphics{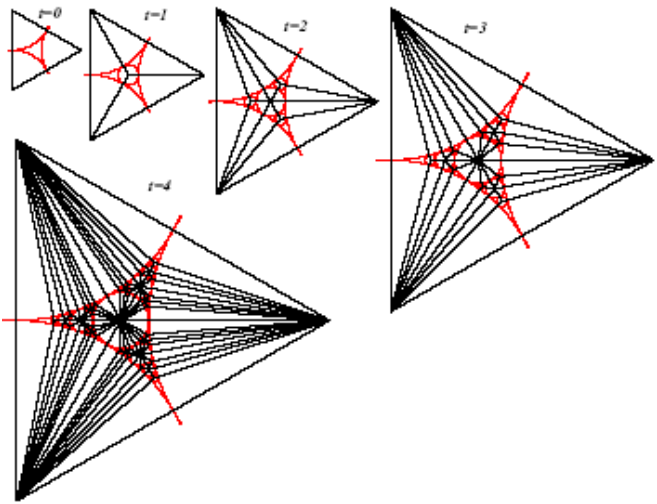}} \caption{The
development of the 2D Apollonian network inside the interstice
between three mutually touching disks, as the number of
generations increase. In each picture, the network is overlaid on
the underlying packing\cite{Doye2004}.}
\end{figure}

Apollonian networks, introduced by Andrade et al
\cite{Andrade2004}, are derived from the problem of space-filling
packing of spheres according to the ancient Greek mathematician
Apollonius of Perga \cite{Boyd1973}. To produce an Apollonian
packing, we start with an initial array of touching disks, the
interstices of which are curvilinear triangles. In the first
generation disks are added inside each interstice in the initial
configuration, such that these disks touch each of the disks
bounding the curvilinear triangles. The positions and radii of
these disks can easily be calculated, and the circle size
distribution follows a power-law with exponent of about 1.3
\cite{Boyd1973}. Of course, these added disks cannot fill all of
the space in the interstices, but instead give rise to three
smaller interstices. In the second generation, further disks are
added inside all of these new interstices, which again touch the
surrounding disks. This process is then repeated for successive
generations. If we denote the number of generations by $t$, where
$t=0$ corresponds to the initial configuration, as $t\rightarrow
\infty$ the space-filling Apollonian packing is obtained as shown
in figure 1. Apollonian packing can be used as a basis of a
network, where each disk is a node in the network and nodes are
connected if the corresponding disks are in contact. We call this
contact an ``Apollonian Network". Figure 2 shows how the network
evolves with the addition of new nodes at each generation. For
each new disk added, three new interstices in the packing are
created, that will be filled in the next generation. Equivalently,
for each new node added, three new triangles are created in the
network, into which nodes will be inserted in the next generation.

Doye et. al. have studied the properties of Apollonian networks
detailedly \cite{Doye2004}, and shown the degree distribution
$p(k)\propto k^{{\rm -\gamma}}$, average length $l\propto (ln
N)^{{\beta}}$, where $\gamma = 1 + \frac{ln 3}{ln 2} \approx
2.585$, $\beta \approx 0.75$ and $N$ is the order \cite{ex1} of
the network, in other words, Apollonian networks are scale-free
and display small-world effect. It is worth remarking that the
clustering coefficient $C$ is close to 0.828, much larger than
that of BA networks, in the limit of large $N$.

Andrade et. al. have also found many peculiar results about some
well-known models upon Apollonian networks, including percolation,
electrical conduction, and magnetic
model\cite{Andrade2004,Andrade2004b}.

\section{Random Apollonian Networks}

\begin{figure}
\scalebox{0.4}[0.4]{\includegraphics{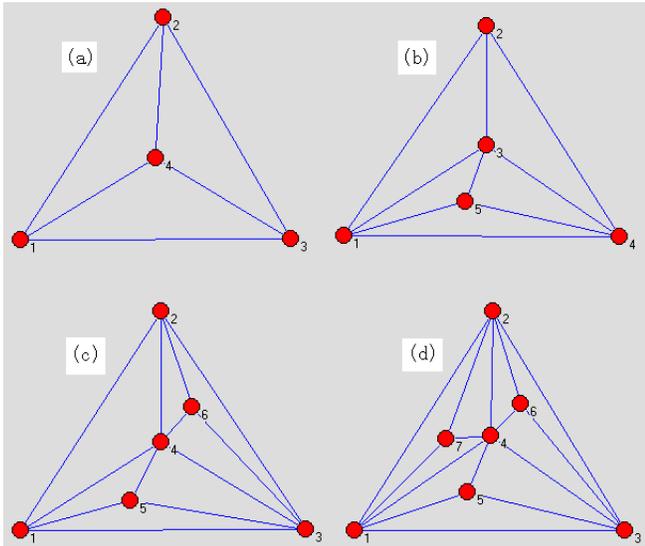}}
\caption{\label{fig:epsart} The sketch maps for the network
growing process. The four figures show a possible growing process
for RAN at time $t=1$(a), $t=2$(b), $t=3$(c) and $t=4$(d). At time
step 1, the 4th node is added to the network and linked to node 1,
2 and 3. Then, at time step 2, the triangle $\triangle 134$ is
selected, the 5th node is added inside this triangle and linked to
node 1, 3 and 4. After that, the triangles $\triangle 234$ and
$\triangle 124$ are selected at time step 3 and 4, respectively.
And node 6 and 7 are added inside these two triangles
respectively. The figure 3d shows a random Apollonian network of
order 7. Keep on the similar iterations, one can get RAN of any
orders as he like.}
\end{figure}

Random Apollonian network starts with a triangle containing three
nodes marked as 1, 2 and 3. Then, at each time step, a triangle is
randomly selected, and a new node is added inside the triangle and
linked to the three vertices of this triangle. The sketch maps for
the network growing process are shown in figure 3.

It is clear that, at time step $t$, our network is of order
$N=t+3$. Using this simple rule, one can get random Apollonian
network of arbitrary order as he like. Note that the randomicity
is involved in our model(that is why we call these networks random
Apollonian networks), the analytic approaches are completely
different from the earlier studies on Apollonian networks.

\section{statistical characteristics of Random Apollonian Networks}

\subsection{The scale-free property}

As we have mentioned above, the degree distribution is one of the
most important statistical characteristics of networks. Since a
majority of real-life networks are scale-free networks, whether
the networks are of power-law degree distribution is a criterion
to judge the validity of the model. In this subsection, we will
give the simulation and analytic results on random Apollonian
networks' degree distribution.

\begin{figure}
\scalebox{0.8}[0.9]{\includegraphics{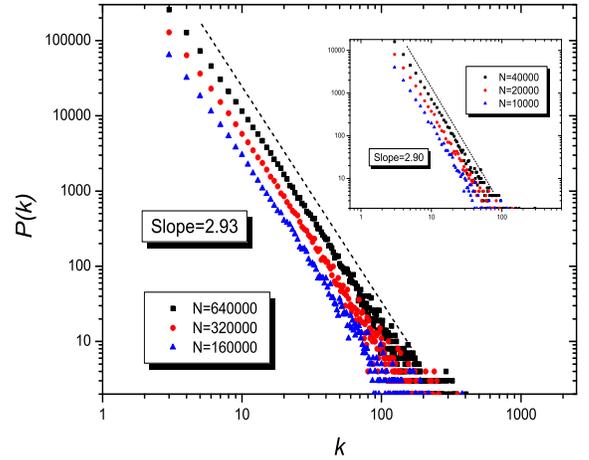}}
\caption{\label{fig:epsart} Degree distribution of RAN, with
$N=640000$(black squares), $N=320000$(red circles) and
$N=160000$(blue triangles). In this figure, $P(k)$ denotes the
number of nodes of degree $k$. The power-law exponent $\gamma$ of
the three probability density function are
$\gamma_{640000}=2.94\pm 0.04$, $\gamma_{320000}=2.92\pm 0.05$ and
$\gamma_{160000}=2.92\pm 0.06$, respectively. The average exponent
of them are 2.93. The inset shows Degree distribution of RAN, with
$N=80000$(black squares), $N=40000$(red circles) and
$N=20000$(blue triangles). The exponents are
$\gamma_{640000}=2.91\pm 0.07$, $\gamma_{320000}=2.90\pm 0.07$ and
$\gamma_{160000}=2.90\pm 0.09$. The mean value is 2.90. The two
dash lines have slope -3.0 for comparison.}
\end{figure}

Note that, after a new node is added to the networks, the number
of triangles increases by 2\cite{ex2}. Therefore, we can
immediately get that when the networks are of order $N$, the
number of triangles are:
\begin{equation}
N_{\triangle}=2(N-3)+1=2N-5
\end{equation}
Let $N_{\triangle}^i$ denote the number of triangles containing
the $i$th node, the probability that a newly added node will link
to the $i$th node is $N_{\triangle}^i/N_{\triangle}$. Apparently,
except the node 1, 2 and 3, $N_{\triangle}^i$ is equal to the
degree of the $i$th node: $N_{\triangle}^i=k_{i}$. Therefore, we
can write down a rate equation\cite{Krapivsky2000} for the degree
distribution. Let $n(N,k)$ be the number of nodes with degree $k$
when $N$ nodes are present, now we add a new node to the network,
$n(N,k)$ evolves according to the following equation:
\begin{equation}
n(N+1,k+1)=n(N,k)\frac{k}{N_{\triangle}}+n(N,k+1)(1-\frac{k+1}{N_{\triangle}})
\end{equation}
When $N$ is sufficient large, $n(N,k)$ can be approximated as
$Np(k)$, where $p(k)$ is the probability density function for the
degree distribution. In terms of $p(k)$, the above equation can be
rewritten as:
\begin{equation}
(N+1)p(k+1)=\frac{Nkp(k)}{N_{\triangle}}+Np(k+1)-\frac{N(k+1)p(k+1)}{N_{\triangle}}
\end{equation}
Using Equ.(1) and the expression $p(k+1)-p(k)=\frac{dp}{dk}$, we
can get the continuous form of Equ.(3):
\begin{equation}
k\frac{dp}{dk}+\frac{3N-5}{N}p(k)=0
\end{equation}
This lead to $p(k)\propto k^{-\gamma}$ with
$\gamma=(3N-5)/N\approx 3$ for large $N$.

In figure 4, we report the degree distribution for
$N=640000,320000,160000,80000,40000$ and 20000. The simulation
results agree very well with the analytic one.

By the way, some readers may think the RAN are almost the same as
BA networks. Indeed, the two ingredients ``growth" and
``preferential attachment" are in common and they have almost the
same power-law exponent, but the further simulation and analytic
results will show that RAN and BA networks are essentially
different in some aspects and RAN may be closer to reality rather
than BA networks.

\subsection{The small-world effect}

\begin{figure}
\scalebox{0.8}[0.9]{\includegraphics{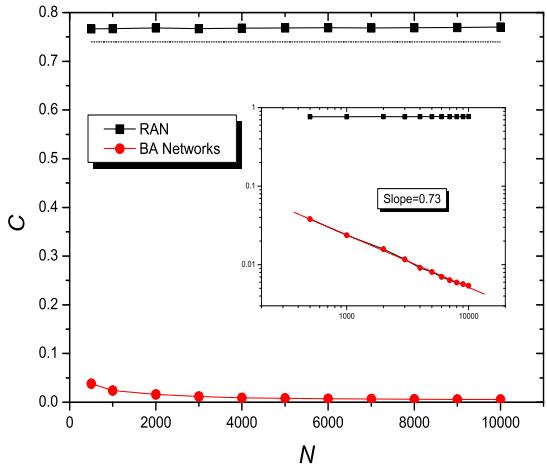}}
\caption{\label{fig:epsart} The clustering coefficient of
RAN(black squares) and BA networks(red circles). In this figure,
one can find that the clustering coefficient of RAN is almost a
constant a little smaller than 0.77, which accord with the
analytic very well. The dash line represents the analytic result
0.74. It is clear that, the clustering coefficient of BA networks
is much smaller than that of RAN. The inset shows that the
clustering coefficient of BA networks decreases with the
increasing of network order, following approximately a power law
$C\sim N^{-0.73}$, which is quite different form the real-life
networks. All the data are obtained by 10 independent
simulations.}
\end{figure}

The small-world effect consists of two properties: large
clustering coefficient and small average distance. In this
subsection, we will give both the simulation and analytic results
about the two properties and prove that RAN are small-world
networks.

At first, let us calculate the clustering coefficient of RAN. As
we mentioned in section 1, for an arbitrary node $x$, the
clustering coefficient $C(x)$ is:
\begin{equation}
C(x)=\frac{2E(x)}{k(x)(k(x)-1)}
\end{equation}
where $E(x)$ is the number of edges among node $x$'s neighbor-set
$A(x)$, and $k(x)=|A(x)|$ is the degree of node $x$. The
clustering coefficient $C$ of the whole network is defined as the
average of $C(x)$ over all nodes.

\begin{figure}
\scalebox{0.8}[0.9]{\includegraphics{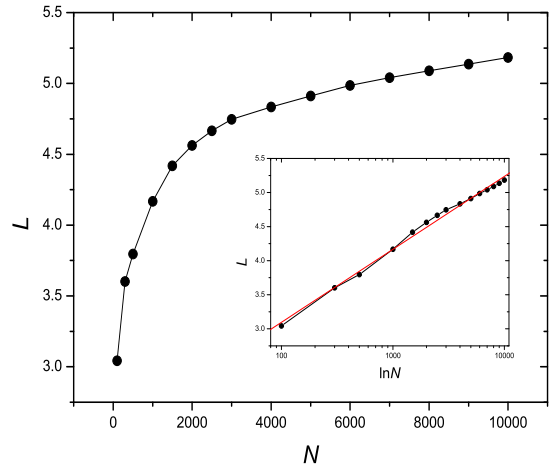}}
\caption{\label{fig:epsart} The dependence between the average
distance $L$ and the order $N$ of RAN. One can see that $L$
increases very slowly as $N$ increases. The inset exhibits the
curve where $L$ is considered as a function of $\texttt{ln}N$,
which is well fitted by a straight line. The curve is above the
fitting line when $N$ is small($1000\leq N\leq 3000$) and under
the fitting line when $N$ is large($N\geq4000$), which indicates
that the increasing tendency of $L$ can be approximated as
$\texttt{ln}N$ and in fact a little slower than $\texttt{ln}N$.
All the data are obtained by 10 independent simulations.}
\end{figure}

By means of theoretic calculation(see {\bf Appendix A} for
details), we obtain the clustering coefficient of RAN with large
order $N$ as:
\begin{equation}
C=\frac{46}{3}-36\texttt{ln}\frac{3}{2}\approx 0.74
\end{equation}

Figure 5 shows the simulation results about the clustering
coefficient of RAN, which agree very well with the analytic one.
It is remarkable that, the clustering coefficient of BA networks
is very small and decreases with the increasing of network order,
following approximately a power law $C\sim
N^{-0.75}$\cite{Albert2002}(In our simulation, it is found to be
about 0.73). The simulation about clustering coefficient of BA
network can also be seen in figure 5. Since the data-flow patterns
show a large amount of clustering in interconnection networks, the
RAN may perform better than BA networks. In addition, the
demonstration exhibits that most real-life networks have large
clustering coefficient no matter how many nodes they have. That is
agree with the case of RAN but conflict with the case of BA
networks, thus RAN may be more appropriate to mimic the reality.

In addition, many real networks including Internet, World Wide
Web, and the actor network, are characterized by the existence of
hierarchical structure\cite{Ravasz2003,Trusina2004,Kim2004b},
which can usually be detected by the negative correlation between
the clustering coefficient and the degree. The BA networks , which
does not possess hierarchical structure, is known to have the
clustering coefficient $C(x)$ of node $x$ independent of its
degree $k(x)$\cite{Ravasz2003}, while the RAN has been shown to
have $C(k)\sim k^{-1}$(see Eq.(A2), in accord with the
observations of many real-life networks\cite{Ravasz2003}.)

In succession, let's discuss the average distance of RAN. By means
of theoretic approximate calculation(see {\bf Appendix B} for
details), we prove that the increasing tendency of $L(N)$ is a
little slower than $\texttt{ln}N$. In figure 6, we report the
simulation results on average distance of RAN, which agree with
the analytic result well.

In respect that the random Apollonian networks are of very large
clustering coefficient and very small average distance, they are
not only the scale-free networks, but also small-world networks.
Since most real-life networks are both scale-free and small-world
networks, RAN may perform better in mimicking reality rather than
WS networks and BA networks.

\section{The planarity of random Apollonian networks}

\begin{figure}
\scalebox{0.8}[0.9]{\includegraphics{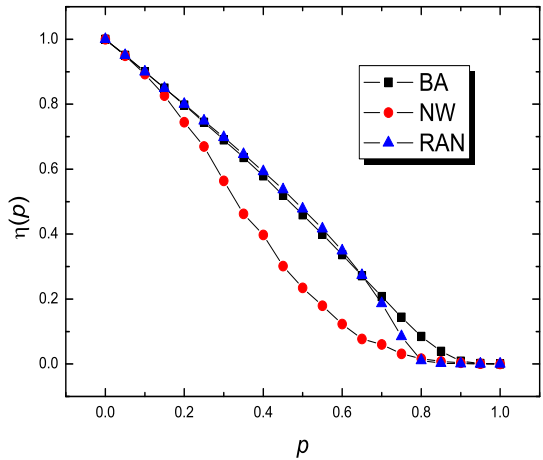}}
\caption{\label{fig:epsart} Random site-percolation transition for
RAN(blue triangles), NW(red circles) and BA(black squares)
networks. Plotted is the fraction of nodes that remain in the
giant component after random breakdown of a fraction $p$ of all
nodes, $\eta(p)$, as a function of $p$. All the networks using for
numerical study are of order $N=10000$ and $\langle k\rangle =6$.
Clearly, the scale-free networks(BA and RAN) are much more
resilient than networks of single-scale(NW), which agrees with the
well-known conclusion\cite{Albert2000a,Cohen2000}. When $p<0.3$,
the performances of RAN and BA networks under random failures are
almost the same; when $0.3<p<0.6$, RAN are little resilient than
BA networks; when $p$ becomes even larger, BA networks get
obviously more resilient than RAN. The critical thresholds of RAN
and BA networks are $p_c^{RAN}\approx 0.85$ and $p_c^{BA}\approx
0.95$, which will approaches 1 as the networks grow in
size\cite{Cohen2000}. All the data are averaged over 100
independent simulations.}
\end{figure}

There are many practical situations in which it is important to
decide whether a given network is planar, and if so, to then find
a planar embedding of the network. For example, a very large scale
integrated(VLSI) designer has to place the cells on printed
circuit boards according to several designing requirements. One of
these requirements is to avoid crossings since they may lead to
undesirable signals. One is, therefore, interested in knowing if a
given electrical network is planar, where the nodes correspond to
electrical cells and the edges to the conductor wires connecting
the cells.

A network is a planar network if it can be drawn in the plane in
such a way that no two edges intersect. Putting it a little more
rigorously, it is possible to represent it by a drawing in the
plane in which the nodes correspond to distinct points and the
edges to simple Jordan curves connecting the points of its
end-points. In this drawing every two curves are either disjoint
or meet only at a common end-point. The above representation of a
graph is said to be a plane networks.

In some places, the networks will perform better when they have
more edges. Therefore, how to add more edges into a network but
keeping it a planar network is a practical and interesting
problem. According to the rule that generates RAN, one can
immediately find that RAN are planar networks. Hereinafter, we
will show that the RAN are maximal planar networks, which are the
planar networks with fixed order who has maximum edges.

If we omit the nodes and edges of a planar networks from the
plane, the remainder falls into connected components, called
faces\cite{Bollobas1998}. Clearly, each plane network has exactly
one unbounded face and each edge is in the boundary of two faces.
If we draw the graph of a convex polyhedron in the plane, then the
faces of the polyhedron clearly correspond to the faces of the
plane networks. This lead to the Euler's polyhedron theorem or
simply Euler's formula\cite{Xu2003,Bond1976,Bollobas1998} that if
a connected plane networks has $n$ nodes, $m$ edges and $f$ faces,
then:
\begin{equation}
n-m+f=2
\end{equation}
Furthermore, denote $f_i$ the number of faces having exactly $i$
edges in their boundaries. Clearly,
\begin{equation}
\sum_if_i=f
\end{equation}
And since each edge is in the boundary of two faces, we have:
\begin{equation}
\sum_iif_i=2m
\end{equation}
Combine Equ.(8) and Equ.(9) and note that each face has at least 3
edges, we have:
\begin{equation}
2m\geq 3\sum_if_i=3f
\end{equation}
Then, using Euler's formula, one can obtain that:
\begin{equation}
m\leq 3n-6
\end{equation}
That is to say the maximum number of edges for a planar network
with order $n$ is $3n-6$. Apparently, the random Apollonian
network with order $N$ has $3N-6$ edges on the beam, thus all the
RAN are maximal planar networks.

\section{Percolation and Epidemic Spreading on Random Apollonian Networks}

As we mentioned above, close to many real-life networks, random
Apollonian networks are both scale-free and small-world.
Therefore, it is worthwhile to investigate the processes taking
place upon RAN and directly compare these results with just
small-world networks(like WS and NW networks) and just scale-free
networks(like BA networks). These comparisons may give us deep
sight into dynamic properties of networks.

In this section, we will exhibit some simulation results on two
deeply studied ones, {\bf percolation} and {\bf epidemic spreading
process}. Since the WS networks may be unconnected, we will use NW
networks as exemplifications of small-world networks.

\subsection{Percolation}

In the year 2000, Albert et. al. have raised the questions of
random failures and intentional attack on
networks\cite{Albert2000a}. Part of these questions can be equally
considered as site-percolation or bond-percolation on
networks\cite{Grassberger1983,Grimmett1989}. In this subsection,
upon RAN, BA and NW networks, we will investigate two kinds of
site-percolation, {\bf random site-percolation}(RSP) and {\bf
preferential site-percolation} (PSP), which correspond to random
failures and intentional attack aiming at nodes, respectively.

When such networks are subject to random breakdowns-a fraction $p$
of the nodes and their incident edges are removed randomly-their
integrity might be compromised: when $p$ exceeds a certain
threshold, $p>p_c$, the network disintegrates into smaller,
disconnected fragments. Below that critical threshold, there still
exists a connected cluster named giant component that spans the
entire system(its size is proportional to that of the entire
system). In figure 7, we report the simulation results for RSP
upon RAN, BA and NW networks. RAN and BA networks are obviously
more resilient than NW networks under random failures that agrees
with the well-known conclusion\cite{Albert2000a,Cohen2000}.
According to the results obtained by Cohen et.
al.\cite{Cohen2000}, RAN and BA networks do not have nontrivial
critical threshold $p_c<1$ in the limit of large network order
$N\rightarrow \infty$. However, for finite network order and large
$p$, RAN are frailer than BA networks.

\begin{figure}
\scalebox{0.8}[0.9]{\includegraphics{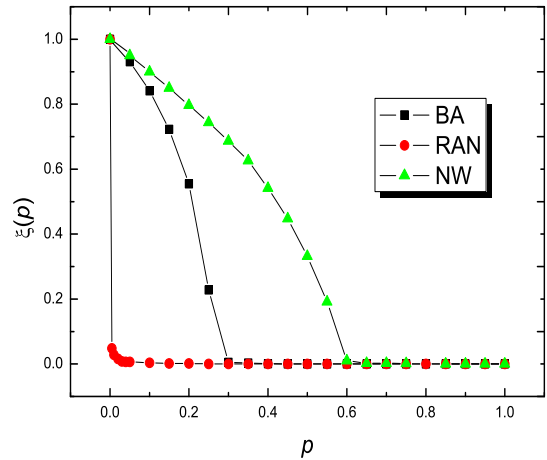}}
\caption{\label{fig:epsart} Preferential site-percolation for
RAN(red circles), BA(black squares) and NW(green triangles)
networks(from left to right) with $N=10000$ and $\langle k\rangle
=6$ fixed. Plotted is the fraction of nodes that remain in the
giant component under intentional attack of a fraction $p$ of all
nodes, $\xi(p)$, as a function of $p$. The critical threshold of
RAN, BA and NW networks are $p_c^{RAN}\approx 0.03$,
$p_c^{BA}\approx 0.3$, and $p_c^{NW}\approx 0.65$, respectively.
All the data are averaged over 100 independent simulations.}
\end{figure}

Figure 8 shows the performances of RAN, BA and NW networks under
intentional attack, which means the removal of nodes is not
random, but rather nodes with the greatest degree are targeted
first. One can find that the scale-free networks are far more
sensitive to sabotage of a small fraction of the nodes, leading
support to the view of Albert and Cohen et.
al.\cite{Albert2000a,Cohen2001}. Although we know the critical
threshold for PSP upon scale-free networks will decay to zero as
the increasing of network order as $\texttt{lim}_{N\rightarrow
\infty}p_c=0$, the notable difference between RAN and BA networks
when $N$ is relatively small surprises us. For very large $N$(as
$N\sim 10^6$ or even larger), the performances of RAN and BA are
almost the same with $p_c<0.03$\cite{Cohen2001}. However, the
susceptivity to order of RAN and BA networks are completely
different. BA networks are very sensitive to network order, for
$N=10000$, its critical threshold is about ten times than the
asymptotic value. And RAN are almost impervious to the change of
network order. Since many real-life networks are of order in the
range $10^3$ to $10^5$, this finding may be valuable in
practicability.

Why RAN and BA networks display completely different susceptivity
to order changing, does it owe to the difference of clustering
structure or hierarchical structure? This problem puzzles us much.
To study the finite-size effect for PSP upon scale-free networks
in detail and to use network model with tunable clustering
coefficient\cite{Holme2002a} may reveal some news, which will be
considered in the future.

\subsection{Epidemic Spreading Process}

Recent studies on epidemic spreading in complex networks indicate
a particular relevance in the case of networks characterized by
various topologies that in many cases present us with new epidemic
propagation scenarios such as the absence of any epidemic
threshold below which the infection cannot initiate a major
outbreak\cite{Pastor-Satorras2001a,Pastor-Satorras2001b}. The new
scenarios are of practical interest in computer virus diffusion
and the spreading of diseases in heterogeneous populations.
However, most previous studies have been focused on the stationary
properties of endemic states or the final prevalence(i. e. the
number of infected individuals) of epidemics. For the sake of
protecting networks and finding optimal strategies for the
deployment of immunization resources, it is of practical
importance to study the dynamical evolution of the outbreaks,
which has been far less investigated before. Barth\'{e}lemy et.
al. reported both the analytic and numerical results of velocity
of epidemic outbreaks in BA networks, which leaves us very short
response time in the deployment of control
measures\cite{Barthelemy2004}. We have studied the same process in
weighted scale-free networks and demonstrated that the larger
dispersion of weight of networks results in slower spreading,
which may be a good news for us\cite{Yan2004}. In this subsection,
we intend to study how the connectivity pattern(i. e. topological
structure) affects the epidemic spreading process in the
outbreaks. Numerical simulations about BA, NW and RAN networks are
drawn, which may give us more comprehensive sight into the
corresponding dynamic behavior.

\begin{figure}
\scalebox{0.8}[0.9]{\includegraphics{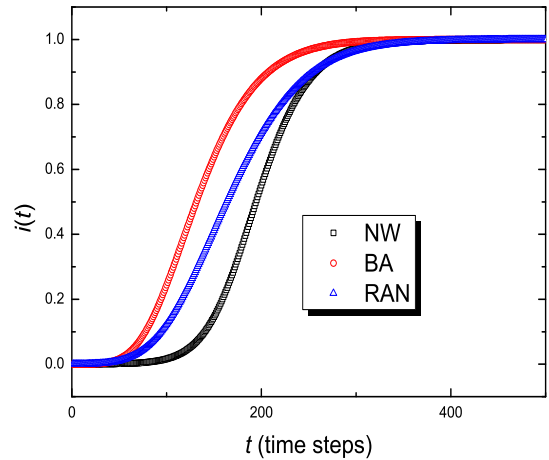}}
\caption{\label{fig:epsart} Average density of infected
individuals versus time in networks with order $N=10000$ and
average degree $\langle k\rangle =6$ fixed. The black, blue and
red curves(from bottom to top) correspond to the case of NW, RAN
and BA networks respectively. The NW networks are of $z=1$ and
$\phi =4\times 10^{-4}$, thus $\langle k\rangle \approx 2z+\phi
N=6$\cite{ex3}(see also the accurate definitions of $z$ and $\phi$
in reference\cite{Newman1999b}). The spreading rate is
$\lambda=0.01$. All the data are averaged over $10^3$ independent
runs.}
\end{figure}

\begin{figure}
\scalebox{0.8}[0.9]{\includegraphics{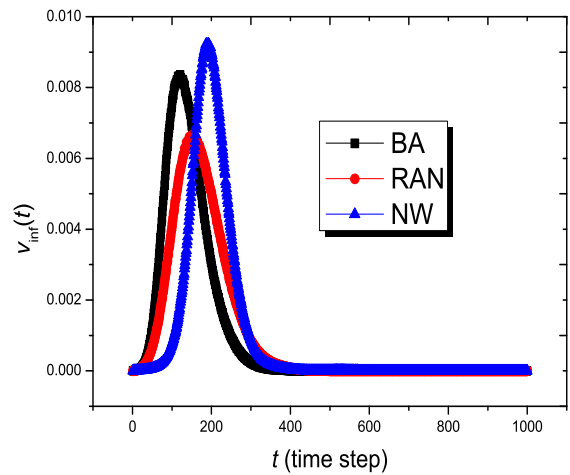}}
\caption{\label{fig:epsart} The spreading velocity vs time. The
network order and spreading rate are the same with figure 9. The
blue, red and black curves correspond to the case of NW, RAN and
BA networks respectively. The spreading velocity reaches a peak
quickly. Before the peak-time, the spreading velocities of the
three kinds of networks satisfy the inequality
$v_{inf}^{BA}>v_{inf}^{RAN}>v_{inf}^{NW}$. All the data are
averaged over $10^3$ independent runs.}
\end{figure}

In order to study the dynamical evolution of epidemic outbreaks,
we shall focus on the susceptible-infected(SI) model in which
individuals can be in two discrete states, either susceptible or
infected\cite{Anderson1992,Murray1993}. Each individual is
represented by a node of the network and the edges are the
connections between individuals along which the infection may
spread. The total population(the network order) $N$ is assumed to
be constant thus if $S(t)$ and $I(t)$ are the number of
susceptible and infected individuals at time $t$, respectively,
then
\begin{equation}
N=S(t)+I(t)
\end{equation}
In the SI model, the infection transmission is defined by the
spreading rate $\lambda$ at which each susceptible individual
acquires the infection from an infected neighbor during one time
step. In this model, infected individuals are assumed to be always
infective, which is an approximation that is useful to describe
early epidemic stages in which no control measures are deployed.
According to Eq.(12), one can easily obtain the probability that
an susceptible individual $x$ will be infected at the present time
step is:
\begin{equation}
\lambda_x(t)=1-\lambda^{\theta(x,t-1)}
\end{equation}
where $\theta(x,t-1)$ denotes the number of infected individuals
at time step $t-1$ in $x$'s neighbor-set $A(x)$.

We start by selecting one node randomly and assume it is infected.
The diseases or computer virus will spread in the networks in
according with the rule of Eq.(13). In figure 9, we plot the
average density over 1000 independent runs in RAN, BA and NW
networks with $N=10000$ and $\langle k\rangle =6$ fixed.
Obviously, all the individuals will be infected in the limit of
long time as $\texttt{lim}_{t\rightarrow \infty}i(t)=1$, where
$i(t)=\frac{I(t)}{N}$ denotes the density of infected individuals
at time step $t$. More over, the simulation results indicate that
the diseases spread more quickly in BA networks than RAN, as well
as in RAN than NW networks. To make the outcome more clear, we
have calculated the diseases spreading velocity, which is defined
as:
\begin{equation}
v_{inf}(t)=\frac{\texttt{d}i(t)}{\texttt{d}t}\approx
\frac{I(t)-I(t-1)}{N}
\end{equation}

Figure 10 shows the spreading velocity vs time in RAN, BA and NW
networks with $N=10000$ and $\langle k\rangle=6$. The spreading
velocity reaches a peak quickly. Before the peak-time, the
spreading velocities of the three kinds of networks satisfy the
inequality:
\begin{equation}
v_{inf}^{BA}>v_{inf}^{RAN}>v_{inf}^{NW}
\end{equation}
The result that diseases spread more quickly in RAN and BA
networks than in NW networks is easy to be understood as the
well-known conclusion: boarder degree distribution will speed up
the epidemic spreading
process\cite{Pastor-Satorras2001a,Pastor-Satorras2001b}.

Why the diseases spread more quickly in BA networks than RAN is a
very interesting question. We argue that the larger clustering
coefficient may slow down the epidemic spreading process
especially in the outbreaks. For an arbitrary edge $e$, containing
two nodes $x$ and $y$, obviously, the distance between $x$ and $y$
is $d(x,y)=1$. Remove the edge $e$ from quondam network, then, the
distance between $x$ and $y$ will increase $d'(x,y)>1$(if the
removal of $e$ makes $x$ and $y$ disconnected, then we set
$d'(x,y)=N$). The quantity $d'(x,y)$ can be considered as edge
$e$'s score $s(e)=d'(x,y)\geq 2$, denoting the number of edges the
diseases must pass through from $x$ to $y$ or form $y$ to $x$ if
they do not pass across $e$. If $s(e)$ is small, then $e$ only
plays a local role in the epidemic spreading process, else when
$s(e)$ is large, $e$ is of global importance. For each edge $e$,
if it does some contribution to clustering coefficient, it must be
contained in at least one triangle and $s(e)=2$. Therefore,
networks of larger clustering coefficient have more {\bf local
edges}. RAN and BA networks are two extreme ones of scale-free
networks. In RAN, all the edges are of score 2; while in BA
networks, almost all the edges are of score larger than 2 because
the clustering coefficient of BA networks will decay to zero
quickly as $N$ increases. Consequently, diseases spread more
quickly in BA networks than RAN.

The above explanation is qualitative and rough, to study the
process upon networks with tunable clustering
coefficient\cite{Holme2002a} may be useful for the present
problem, which will be one of the future works.

\section{Conclusions}
In conclusion, in respect that the random Apollonian networks are
of very large clustering coefficient and very small average
distance, they are not only the scale-free networks, but also
small-world networks. Since most real-life networks are both
scale-free and small-world networks, RAN may perform better in
mimicking reality rather than WS networks and BA networks. In
addition, RAN possess hierarchical structure that in accord with
the observations of many real networks and we propose an analytic
approach to calculate clustering coefficient. Since in the earlier
studies, only few analytic results about clustering coefficient of
networks with randomicity are
reported\cite{Barrat2000,Newman2002,Ravasz2003,Szabo2003}, we
believe that our work may enlightened readers on this subject.

Further more, we briefly introduce the conception of planar
network(it is also called ``planar graph" in mathematical
language), and prove that RAN are maximal planar networks, which
are of particular practicability for layout of printed circuits
and so on. Although whether a network is planar or not is a
natural and important question that attracts much attention for
mathematicians, it seems not interesting for physicists and almost
no pertinent results are reported in the earlier studies on
complex networks. But in fact, many real-life networks are planar
networks by reason of technical or natural requirements, such as
layout of printed circuits, river networks upon the earth's
surface, vas networks clinging to cutis, and so forth. Since the
planar networks have some graceful characteristics that can not be
found in non-planar ones, researchers ought to pay more attention
to networks' planarity. We wish our abecedarian work to stimulate
physicists thinking more of planarity.

The percolation and epidemic spreading process are also studies
and the comparison between RAN and BA as well as NW networks are
shown. In percolation model, we find that, when the network order
$N$ is relatively small(as $N\sim 10^4$), the performance of RAN
under intentional attack is not sensitive to $N$, while that of BA
networks is much affected by $N$. In epidemic spreading process,
the diseases spread slower in RAN than BA networks during the
outbreaks, indicating that the large clustering coefficient may
slower the spreading velocity especially in the outbreaks. We give
some qualitative explanation about how the clustering structure
affect the spreading process in the outbreaks, but Why RAN and BA
networks display completely different susceptivity to order
changing is even a problem puzzling us. Those simulation results
suggest that the clustering structure may affect the dynamical
behavior upon networks much. Since many real-life networks are of
great clustering coefficient, to study the process upon networks
with tunable clustering coefficient\cite{Holme2002a} is
significant.

\begin{acknowledgments}

This work has been partially supported by the State Key
Development Programme of Basic Research (973 Project) of China,
the National Natural Science Foundation of China under Grant No.
70471033, 10472116 and No.70271070, the Specialized Research Fund
for the Doctoral Program of Higher Education (SRFDP
No.20020358009), and the Foundation for Graduate Students of
University of Science and Technology of China under Grant No.
USTC-SS-0501.

\end{acknowledgments}

\appendix

\section{More details for calculation of clustering coefficient}

At first, let us consider the clustering coefficient $C(x)$ of an
arbitrary node $x$ except the nodes 1, 2 and 3. At the very time
when node $x$ is added to the network, it is of degree 3 and
$E(x)=3$. After, if the degree of node $x$ increases by one(i.e. a
new node is added to be a neighbor of $x$) at some time step, then
$E(x)$ will increases by two since the newly added node will link
to two of the neighbors of node $x$. Therefore, we can write down
the expression of $E(x)$ in terms of $k(x)$:
\begin{equation}
E(x)=3+2(k(x)-3)=2k(x)-3
\end{equation}
Using Equ.(5), we can get the clustering coefficient of node $x$
as:
\begin{equation}
C(x)=\frac{2(2k(x)-3)}{k(x)(k(x)-1)}
\end{equation}
Consequently, we have:
\begin{equation}
C=\frac{2}{N}\sum_{i=1}^{N}\frac{2k_i-3}{k_i(k_i-1)}=\frac{2}{N}\sum_{i=1}^{N}(\frac{3}{k_i}-\frac{1}{k_i-1})
\end{equation}
where $k_i$ denotes the degree of the $i$th node. Rewrite
$\sum_{i=1}^{N}f(k_i)$ in continuous form:
\begin{equation}
\sum_{i=1}^{N}f(k_i)=\int_{k_{min}}^{k_{max}}Np(k)f(k)dk
\end{equation}
where $k_{min}$ and $k_{max}$ denote the minimal and maximal
degree in RAN, respectively. Then, Equ.(A3) can be rewritten as:
\begin{equation}
C=6\int_{k_{min}}^{k_{max}}\frac{p(k)}{k}dk-2\int_{k_{min}}^{k_{max}}\frac{p(k)}{k-1}dk
\end{equation}
In section 4, we have proved that $p(k)=\alpha k^{-\gamma}$ with
$\gamma =3$ and $\alpha$ a constant, thus one can write down the
expression that:
\begin{equation}
C=6\alpha \int_{k_{min}}^{k_{max}}k^{-4}dk-2\alpha
\int_{k_{min}}^{k_{max}} \frac{1}{k^3(k-1)}dk
\end{equation}
Note that $1/k^3(k-1)=1/(k-1)-1/k-1/k^2-1/k^3$, one can
immediately obtain the value of $C$ as:
\begin{widetext}
\begin{equation}
C=2\alpha
(\frac{1}{k_{min}^3}+\frac{1}{2k_{min}^2}-\frac{1}{k_{min}}-\frac{1}{k_{max}}-\frac{1}{2k_{max}^2}-\frac{1}{k_{max}^3}-\texttt{ln}\frac{k_{min}(k_{max}-1)}{k_{max}(k_{min}-1)})
\end{equation}
\end{widetext}
It is clear that $k_{min}=3$ and for sufficient large $N$,
$k_{max}\gg k_{min}$, and $\alpha$ satisfy the normalization
equation:
\begin{equation}
\int_{k_{min}}^{k_{max}}p(k)dk=1
\end{equation}
Therefore, $\alpha =18$ and
$C=\frac{46}{3}-36\texttt{ln}\frac{3}{2}\approx 0.74$.

\section{The average distance}

At first, let's prove an interesting property about the shortest
paths in RAN. Marked each node according to the time when the node
is added to the network(see figure 3), then we have the following
lemma:

\textbf{Lemma:} for any two nodes $i$ and $j$, each shortest path
from $i$ to $j$($SP_{ij}$) does not pass through any nodes $k$
satisfying that $k>max\{i,j\}$.\\
\emph{Proof.} Using nodes' sequence $$i\rightarrow x_1\rightarrow
x_2 \rightarrow \cdots \rightarrow x_n \rightarrow j$$ to denote
the shortest path from $i$ to $j$ of length $n+1$. Obviously,
$n=0$ is the trivial case. Suppose that $n>0$ and
$x_k=max\{x_1,x_2,\cdots ,x_n\}$, if $\forall 1\leq k\leq n,
x_k<max\{i,j\}$, then the proposition is true.

In succession, we prove that the case $x_k>max\{i,j\}$ would not
come forth. Suppose that the triangle $\triangle y_1y_2y_3$ is
selected when node $x_k$ is added. Since $x_k>max\{i,j\}$, neither
$i$ nor $j$ is inside the triangle $\triangle y_1y_2y_3$. Hence
the path from $i$ to $j$ passing through $x_k$ must enter into and
leave $\triangle y_1y_2y_3$. We may assume that the path enter
into $\triangle y_1y_2y_3$ by node $y_1$ and leave from node
$y_2$, then there exists a subpath of $SP_{ij}$ from $y_1$ to
$y_2$ passing through $x_k$, which is apparently longer than the
direct path $y_1\rightarrow y_2$. Hence if $SP_{ij}$ is the
shortest path, the youngest node must be either $i$ or $j$.
$\blacksquare$

Using symbol $d(i,j)$ to represent the distance between $i$ and
$j$, the average distance of RAN with order $N$, denoted by
$L(N)$, is defined as:
\begin{equation}
L(N)=\frac{2\sigma (N)}{N(N-1)}
\end{equation}
where the total distance is:
\begin{equation}
\sigma (N)=\sum_{1\leq i<j\leq N}d(i,j)
\end{equation}

According to the lemma, newly added node will not affect the
distance between old nodes. Hence we have:
\begin{equation}
\sigma(N+1)=\sigma(N)+\sum_{i=1}^N d(i,N+1)
\end{equation}
Assume that the node $N+1$ is added into the triangle $\triangle
y_1y_2y_3$, then the Equ.(B3) can be rewritten as:
\begin{equation}
\sigma(N+1)=\sigma(N)+\sum_{i=1}^N
(D(i,y)+1)=\sigma(N)+N+\sum_{i=1}^N D(i,y)
\end{equation}
where $D(i,y)=min\{d(i,y_1),d(i,y_2),d(i,y_3)\}$. Constrict
$\triangle y_1y_2y_3$ continuously into a single node $y$, then we
have $D(i,y)=d(i,y)$. Since $d(y_1,y)=d(y_2,y)=d(y_3,y)=0$, the
Equ.(B4) can be rewritten as:
\begin{equation}
\sigma(N+1)=\sigma(N)+N+\sum_{i\in \Gamma} d(i,y)
\end{equation}
where $\Gamma =\{1,2,\cdots ,N\}-\{y_1,y_2,y_3\}$ is a node set
with cardinality $N-3$.

The sum $\sum_{i\in \Gamma} d(i,y)$ can be considered as the total
distance from one node $y$ to all the other nodes in RAN with
order $N-2$. In a rough version, the sum $\sum_{i\in \Gamma}
d(i,y)$ is approximated in terms of $L(N-2)$:
\begin{equation}
\sum_{i\in \Gamma} d(i,y)\approx (N-3)L(N-2)
\end{equation}
Note that, the average distance $L(N)$ increases monotonously with
$N$, it is clear that:
\begin{equation}
(N-3)L(N-2)=\frac{2\sigma(N-2)}{n-2}<\frac{2\sigma(N)}{N}
\end{equation}
Combining (B5), (B6) and (B7), one can obtain the inequation:
\begin{equation}
\sigma(N+1)<\sigma(N)+N+\frac{2\sigma(N)}{N}
\end{equation}

If (B8) is not an inequation but an equation, then the increasing
tendency of $\sigma(N)$ is determined by the equation:
\begin{equation}
\frac{d\sigma(N)}{dN}=N+\frac{2\sigma(N)}{N}
\end{equation}
This equation leads to
\begin{equation}
\sigma(N)=N^2\texttt{ln}N+H
\end{equation}
where $H$ is a constant. As $\sigma(N)\sim N^2L(N)$, we have
$L(N)\sim \texttt{ln}N$. Which should be pay attention to, since
(B8) is an inequation indeed, the precise increasing tendency of
$L$ may be a little slower than $\texttt{ln}N$.


\begin{thebibliography}{Albert2002}
\bibitem{ex1} In this paper, the terminology ``order" means the number of nodes in
network, which is also called network size in some other papers.
\bibitem{ex2} The selected triangle is destroyed
and replaced by three new triangles, thus the number of triangles
increases by 2.
\end{thebibliography}
\end{document}